\documentclass[11pt,sort&compress]{elsarticle}
\usepackage{fullpage}
\usepackage{amsmath}
\usepackage{amssymb}
\usepackage{hyperref}
\usepackage{subfigure}
\usepackage{tikz}
\usetikzlibrary{arrows,shapes,positioning,shadows,trees}
\usepackage{xspace}

\newcommand{\gameA}{{\small \textbf{Game~A}}\xspace}
\newcommand{\gameB}{{\small \textbf{Game~B}}\xspace}
\newcommand{\actionA}{{\small \textbf{Action~A}}\xspace}
\newcommand{\branchOne}{{\small \textbf{Branch~1}}\xspace}
\newcommand{\branchTwo}{{\small \textbf{Branch~2}}\xspace}

\usepackage{tabularx}
\usepackage{booktabs}
\newcolumntype{Y}{>{\centering\arraybackslash}X}    % Y column type provides horizontal centering for X types.
\newcolumntype{L}[1]{>{\raggedright\let\newline\\\arraybackslash\hspace{0pt}}m{#1}}
\newcolumntype{C}[1]{>{\centering\let\newline\\\arraybackslash\hspace{0pt}}m{#1}}
\newcolumntype{R}[1]{>{\raggedleft\let\newline\\\arraybackslash\hspace{0pt}}m{#1}}
\newcolumntype{P}[1]{>{\raggedright\arraybackslash}p{#1}}

\makeatletter
\def\ps@pprintTitle{%
   \let\@oddhead\@empty
   \let\@evenhead\@empty
   \let\@oddfoot\@empty
   \let\@evenfoot\@oddfoot
}
\makeatother

\begin{document}
    
\begin{frontmatter}

\title{Passive network evolution promotes group welfare in complex networks}

\author[sme]{Ye Ye}
\author[smse]{Xiao Rong Hang}
\author[smc]{Jin Ming Koh}
\author[iitis]{Jaros{\l}aw Adam Miszczak}
\author[smc,ec]{Kang Hao Cheong}
\author[smse]{Neng-gang Xie\corref{cor1}}
\ead{xienenggang@aliyun.com}

\address[sme]{School of Mechanical Engineering, Anhui University of Technology, 
Anhui Ma’anshan, 243002, China}
\address[smse]{School of Management Science and Engineering, Anhui University of Technology, Anhui Ma’anshan, 243002, China}
\address[smc]{Science and Math Cluster, Singapore University of Technology and Design (SUTD), 8 Somapah Road, S487372, Singapore}
\address[ec]{SUTD-Massachusetts Institute of Technology International Design Center, 8 Somapah Road, S487372, Singapore}
\address[iitis]{Institute of Theoretical and Applied Informatics, Polish Academy 
of Sciences,  Ba{\l}tycka 5, 44-100 Gliwice, Poland}

\cortext[cor1]{Corresponding author}

\begin{abstract}
The Parrondo's paradox is a counter-intuitive phenomenon in which individually losing strategies, canonically termed \gameA and \gameB, are combined to produce winning outcomes. In this paper, a co-evolution of game dynamics and network structure is adopted to study adaptability and survivability in multi-agent dynamics. The model includes \actionA, representing a rewiring process on the network, and a two-branch \gameB, representing redistributive interactions between agents. Simulation results indicate that stochastically mixing \actionA and \gameB can produce enhanced, and even winning outcomes, despite \gameB being individually losing. In other words, a \textit{Parrondo-type} paradox can be achieved, but unlike canonical variants, the source of agitation is provided by passive network evolution instead of an active second game. The underlying paradoxical mechanism is analyzed, revealing that the rewiring process drives a topology shift from initial regular lattices towards scale-free characteristics, and enables exploitative behavior that grants enhanced access to the favourable branch of \gameB.
\end{abstract}


%\maketitle

\end{frontmatter}

%%%%%%%%%%%%%%%%%%%%%%%%%%%%%%%%%%%%%%%%%%%%%%%%%%%%%%%%%%%%%%%%%%%%%%%%%%%%%%%%
\section{Introduction}
%%%%%%%%%%%%%%%%%%%%%%%%%%%%%%%%%%%%%%%%%%%%%%%%%%%%%%%%%%%%%%%%%%%%%%%%%%%%%%%%

The mechanistic removal and rewiring of connections between nodes promote structural evolution of complex networks \cite{pacheco2006coevolution}. When complex networks are utilized as a modelling basis for multi-agent systems, the topological features resulting from these processes, such as emergent scale-free, small-world and community structural properties, can be expected to carry significant impact on system dynamical behavior. In particular, it is reasonable to inquire whether network evolution may promote group adaptability and survivability, such that formally sub-optimal behavioral patterns yield improved outcomes---in other words, whether \textit{Parrondo-type paradoxes} can be achieved. Investigating the possibility and feasibility conditions of these paradoxes, and more broadly the effects of network evolution on multi-agent game dynamics, is the focus of this paper.

A characterizing property of canonical Parrondo-type game pairs is the achievement of \textit{winning} outcomes through the combination of individually \textit{losing} strategies \cite{harmer1999game}. An \textit{agitation-ratcheting} mechanism underlies the exhibited paradoxical behavior, typically realized through asymmetry in the branching structure of \gameB, such that some branches are favourable (of a higher winning probability), and some are unfavourable (of a higher losing probability). This asymmetric structure forms a \textit{ratchet}. On the other hand, \gameA serves an \textit{agitating} role, to the effect of perturbing the capital of the player. The agitation from \gameA can lead to increased likelihood of landing in favourable branches when \gameB is subsequently invoked, thus manifesting a ratcheting mechanism and enabling the characteristic paradoxical winning outcomes. There have been many examples of such counter-intuitive dynamics studied to date, for instance, in ecological populations \cite{tan2017nomadic,cheong2016paradoxical,koh2018nomadic,cheong2018time,cheong2018do,cheong2017multicellular,koh2019new,cheong2019review}, population genetics \cite{wolf2005diversity,reed2007two,masuda2004subcritical}, physical quantum systems \cite{abbott2010asymmetry,flitney2003quantum,kovsik2007quantum,pawela2013cooperative,miszczak2014general,flitney2002quantum}, reliability theory \cite{di2006parrondo}, system design optimization \cite{koh2018automated, cheong2019hybrid}, and the Allison mixture in information thermodynamics \cite{cheong2017allison}.

A greatly illustrative example is that of the \textit{catfish effect}, deriving from observations of Norwegian fishermen that the forced cohabitation of captured sardines and predatory catfishes can, in fact, be beneficial to sardine yield. The introduction of catfishes into sardine holding tanks stimulates sardine movement, keeping the sardines alive and therefore fresh for a longer duration. A motile state is hence interpretable as \textit{winning}, whereas inactivity or death reflects a \textit{losing} state. Modelling this as a game pair, \gameA then refers to the situation where sardines are in proximity to an active catfish, in which death due to predation (losing) is more likely than survival (winning). \gameA is hence losing. A two-branch \gameB, on the other hand, models sardine behavior when the catfish is absent or otherwise inactive at predation. In \branchOne, the sardines are surrounded by more swimming (winning) individuals, and the probability of winning is large as the surrounding sardines sustain a mixing of water and reduces the risk of hypoxia. In \branchTwo, the sardines are surrounded by more inactive (losing) individuals, and the probability of losing is large due to hypoxia. The tendency of sardines to reduce energy expenditure means that, in the absence of catfishes, the sardine group is likely to invoke \branchTwo, the unfavourable branch; however, with catfish present (\gameA), the group is likely to swim to avoid predation, invoking the favourable \branchOne. This leads to the observed catfish effect.

The catfish effect demonstrates the keystone importance of \textit{agitation} in realizing paradoxical outcomes---indeed, if the agitating \gameA is absent, the ratcheting mechanism cannot at all take effect. In similar vein, the diversity of Parrondo-type paradoxes examined in literature commonly rely on agitation provided by an active game, analogous to \gameA, to drive ratcheting in an accompanying game. In the current paper, we propose the use of \textit{passive} network evolution mechanisms---with no dependence nor influence on agent capital---to provide agitative effects, in replacement of the active \gameA. Existing research relevant to such a construction is notably limited. A previous study had indicated plausible agitative contributions from network rewiring on one-dimensional line and two-dimensional lattice topologies \cite{ye2013multi}, but the results were constrained to regular networks with isomorphic neighborhoods for all nodes. To maintain network regularity, the rewiring process was also constrained to maintain node degrees. These restrictions had enabled simplicity in the mathematics of \gameB but compromises applicability, for networks useful in modelling real-world phenomena are typically non-regular, and rewiring processes do not necessarily preserve degree distributions. 

The shift from the canonical active perturbative sources to a passive one is a key contribution of this study. We adopt an \actionA + \gameB framework, modelling passive network evolution and multi-agent redistributive capital dynamics respectively, to investigate the possibility of emergent Parrondo-type paradoxes. Our results indicate that network structure evolution can indeed enhance group adaptability and survivability, in some cases with sufficiently strong effect as to turn formerly losing behaviors into winning ones. The mechanism underlying these paradoxical outcomes is also elucidated, with novel analyses on evolutionary topological trends towards scale-free characteristics.

\section{Methods}\label{sec:model}

Here, we present a general model comprising a passive rewiring-driven network evolution mechanism, \actionA, and a game capturing multi-agent capital dynamics, \gameB. The structure of the adopted \gameB is shown in  Figure \ref{fig:gameb}, adapted from a previous study \cite{ye2013parrondo}. For each individual, represented as nodes in the network, the two branches of \gameB is selected on a basis of comparison between the capital of the individual $C_i$ and the mean capital of its neighbors $C_i^\dagger$. This branching structure reflects a degree of correlation (or anticorrelation) between the welfare of individuals and their environments, as is indeed expected in the real-world. In \branchOne, played when $C_i\leq{C_i^\dagger}$, the winning probability is $p_1$, and in \branchTwo, played when $C_i>{C_i^\dagger}$, the winning probability is $p_2$. Accordingly, the losing probabilities are $(1-p_1)$ and $(1-p_2)$ respectively. We take winning and losing outcomes at each game round to respectively result in a unit increment and decrement in the capital of the individual. It is notable that the current construction is general to node degree, and is hence compatible with rewiring processes that do not preserve degree distributions, unlike prior models \cite{ye2013parrondo, ye2016effects}.

\tikzset{
    game/.style = {draw, text width=1.6cm, font=\sffamily, rectangle, rounded 
        corners=1pt, thin, align=center},
    outcome/.style = {align=center, text width=2.2em},
    arrow/.style={->,shorten >=5pt,shorten <=5pt,>=stealth}
}

\begin{figure}[ht!]
	\begin{center}
\begin{tikzpicture}
\node[game] (gameb) {\gameB};
\node[game] (branch1) [below left=1.5cm and 1cm of gameb] {\branchOne};
\node[game] (branch2) [below right=1.5cm and 1cm of gameb]  {\branchTwo};

\node[outcome] (wb1) [below left=1.5cm and 0.2cm of branch1] {win};
\node[outcome] (lb1) [below right=1.5cm and 0.2cm of branch1]  {lose};
\node[outcome] (wb2) [below left=1.5cm and 0.2cm of branch2] {win};
\node[outcome] (lb2) [below right=1.5cm and 0.2cm of branch2]  {lose};

\draw[arrow] (gameb) -- (branch1) node[midway, above, xshift=-1.2cm] 
{$C_i \leq \sum_{j\sim i}C_j/K_i$}; 
\draw[arrow] (gameb) -- (branch2) node[midway, above, xshift=1.2cm] 
{$C_i > \sum_{j\sim i}C_j/K_i$}; 

\draw[arrow] (branch1) -- (wb1) node[midway, above, xshift=-0.5cm] 
{$p_1$};
\draw[arrow] (branch1) -- (lb1) node[midway, above, xshift=0.5cm] 
{$1-p_1$};

\draw[arrow] (branch2) -- (wb2) node[midway, above, xshift=-0.5cm] 
{$p_2$}; 
\draw[arrow] (branch2) -- (lb2) node[midway, above, xshift=0.5cm] 
{$1-p_2$};

\end{tikzpicture}
\end{center}
\caption{Construction of \textbf{Game B} on arbitrary complex 
networks \cite{ye2013parrondo}. The summation is taken over the neighbours $j$ 
of node $i$, $j\sim i$. $K_i$ denotes the degree (number of neighbours) of node $i$.}
\label{fig:gameb}
\end{figure}
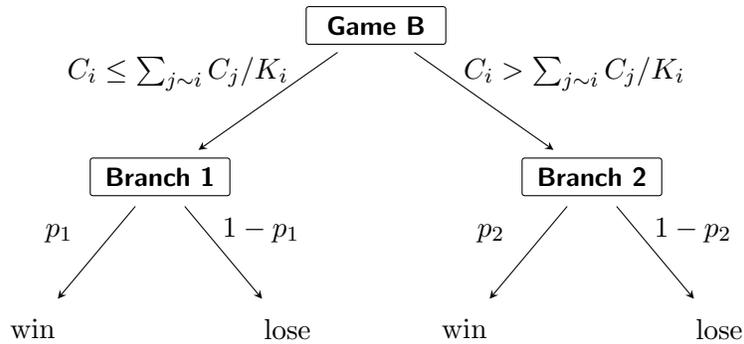

\begin{figure}[!t]
	\centering
		\includegraphics[width=0.7\linewidth]{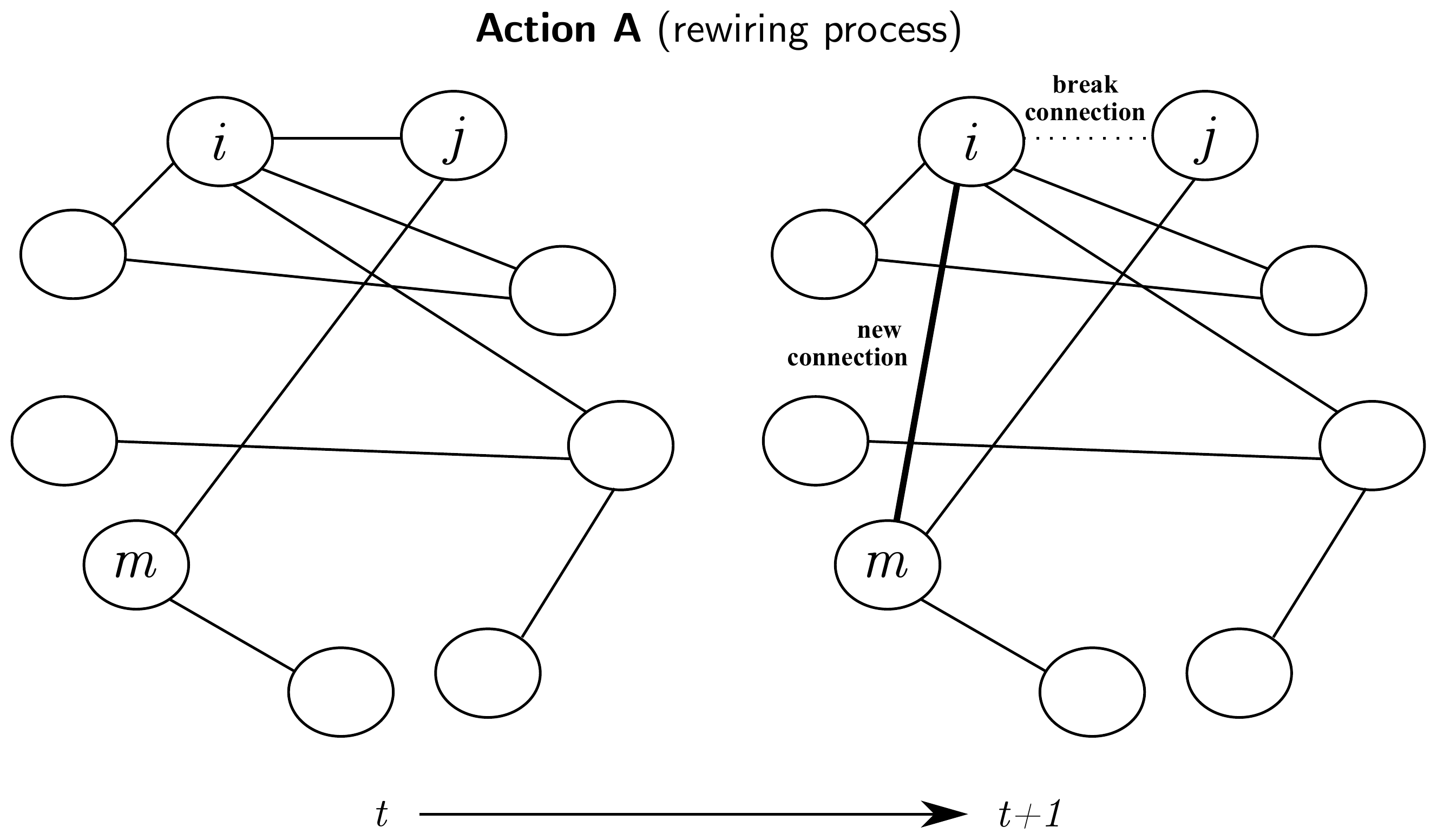}
		\caption{Illustration of a single stpe in the random rewiring mechanism of Action A. New connection (thick line) is created between $I$ and $m$ and connection between $i$ and $j$ is removed (dotted line). For the sake of clarity only $N=9$  nodes are shown.}
	\label{fig:actionA}
\end{figure}

As discussed in the introduction, to enhance the relevance of the network evolution model to real-world systems, the adopted rewiring mechanism is vastly different from previously developed variants \cite{ye2013multi}. There is no constraint in the current rewiring mechanism that the number of neighbors of all nodes be preserved throughout the evolution process. The adopted rewiring mechanism, \actionA, can be described as follows:
\begin{description}
    \item[Step 1:] Node $i$ is selected randomly from the network. If the 
    degree of all neighbors of node $i$ is 1, then node $i$ is re-selected.
    \item[Step 2:] Node $j$ is randomly selected from the neighbors of node 
    $i$. If the degree of node $j$ is 1, then node $j$ is re-selected. If all the neighbors of node $j$ are connected to node $i$ (or is node $i$ itself), then node $j$ is re-selected. In the case that it is not possible to select node $j$ satisfying these properties, then return to \textbf{Step 1} and re-select node $i$.
    \item[Step 3:] Node $m$ is randomly selected from the neighbors of the 
    node $j$. If nodes $m$ and $i$ are identical or are connected, then node $m$ is re-selected.
    \item[Step 4:] The connection between node $i$ and node $j$ is broken, and a connection between nodes $m$ and $i$ is formed.
\end{description}

An illustration of the mechanism of \actionA is shown in Figure \ref{fig:actionA}.\actionA and \gameB can occur individually in the network over a duration of time, such that at each time step $t$, only \actionA or \gameB consistently occurs. Alternatively, a stochastically mixed \actionA + \gameB can be implemented, such that at each time step, either \actionA or \gameB is selected to occur on a random basis. We denote the total population gains of game $B$ and the stochastically mixed \actionA + \gameB as $W^{(B)}(t)$ and $W^{(A+B)}(t)$ respectively. Here, $W(t)=\sum_{i=1}^N \left(C_i(t) - C_0\right)$, where $C_0$ is the initial capital, taken to be equivalent for all individuals, $C_i(t)$ is the capital of individual $i$ at time $t$, and $N$ is the number of individuals (population size) in the network. In ecological or social-dynamical systems, for instance, the capital can be taken to refer to the wealth or welfare of each individual, and the population gain is then a measure of the growth of wealth or welfare of the entire group.

A \textit{Parrondo-type paradox} then occurs when the stochastically mixed \actionA + \gameB produces winning results, despite \gameB being individually losing. Such a paradox is characterized by the condition
\begin{eqnarray}
W^{(A+B)}(t) \geq 0, \qquad W^{(B)}(t) < 0.
\end{eqnarray}

\section{Results \& Discussion}\label{sec:simulations}

Key simulation results and accompanying analyses are presented in this section, starting with observations on the occurrence of a Parrondo-type paradox in the model (Section \ref{sec:simulations/paradox}), network evolution trends (Section \ref{sec:simulations/structure}), elucidation of the underlying mechanism responsible for the paradoxical outcomes (Sections \ref{sec:simulations/macro} and \ref{sec:simulations/micro}), and the presence of a critical rewiring duration after which positive population gains can be maintained even without rewiring (Section \ref{sec:simulations/agitation}). For all simulations, an initial two-dimensional lattice network of $N=60\times 60$ nodes was adopted. The initial capital of all individuals is taken to be $C_0 = 500$, and \actionA or \gameB are played with identical probabilities of $1/2$ on each round for stochastic mixing, over a period of $T = 1.8\times 10^5$ rounds.

\subsection{Occurrence of Parrondo-like paradoxes}\label{sec:simulations/paradox}

\begin{figure*}[!t]
    \centering
    \begin{tabularx}{0.6\textwidth}{@{} m{0.5\textwidth} @{}}
        \centering\includegraphics[width=0.5\textwidth]{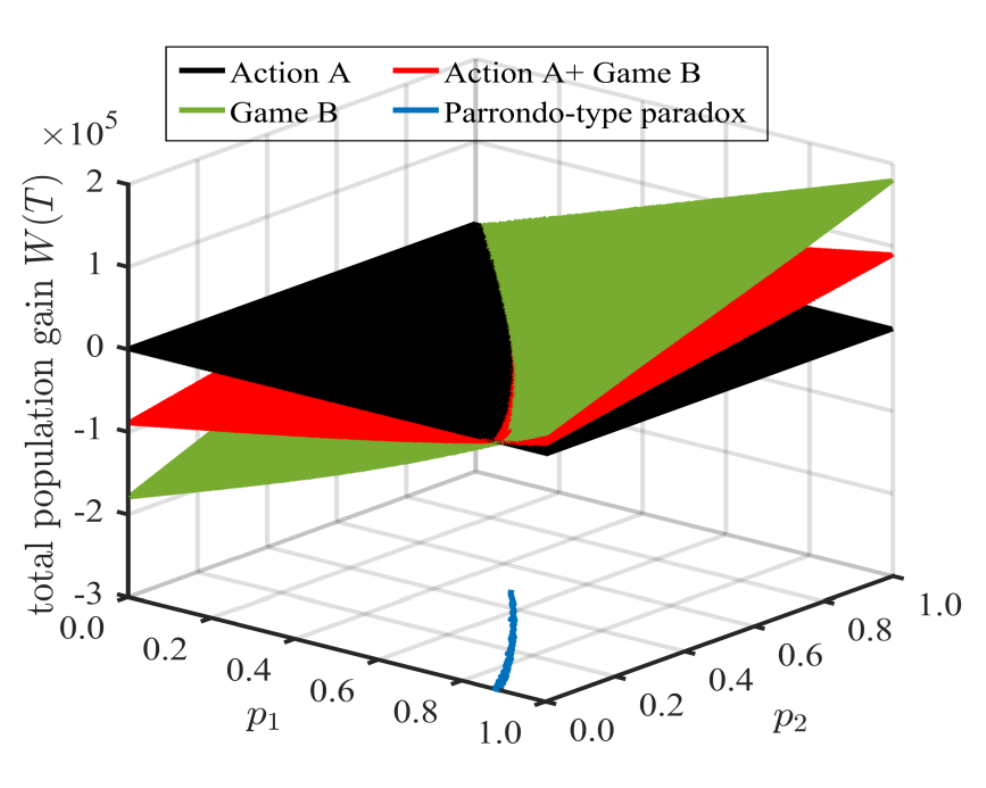}
        \cr\\[-1.5em]
        \centering{\hspace{12pt}\scriptsize(a)}
    \end{tabularx}
    \begin{tabularx}{\textwidth}{@{} m{0.5\textwidth} m{0.5\textwidth} @{}}
        \centering\includegraphics[width=0.5\textwidth]{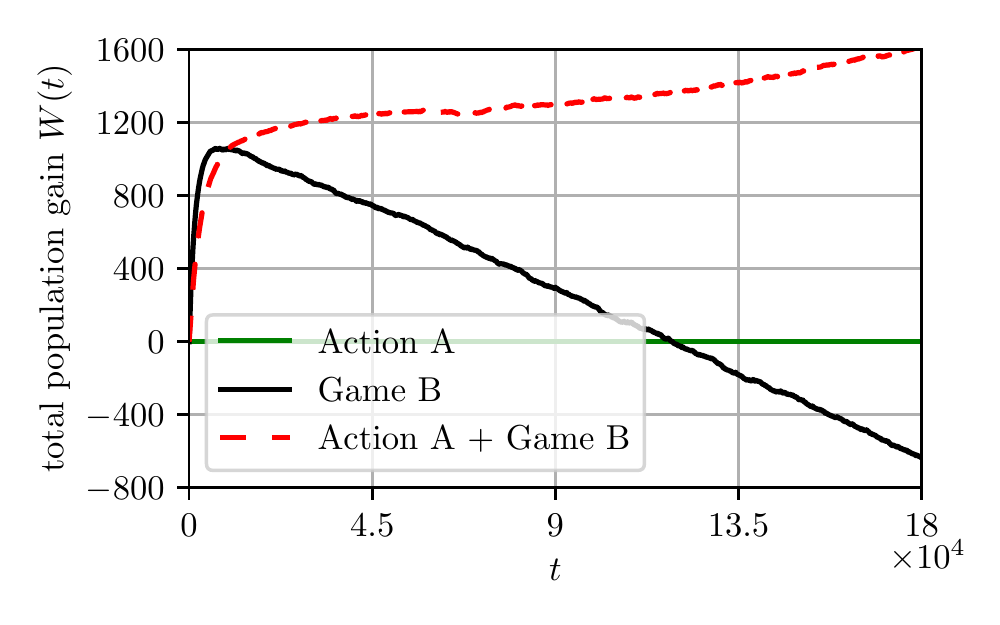}
        &
        \centering\includegraphics[width=0.5\textwidth]{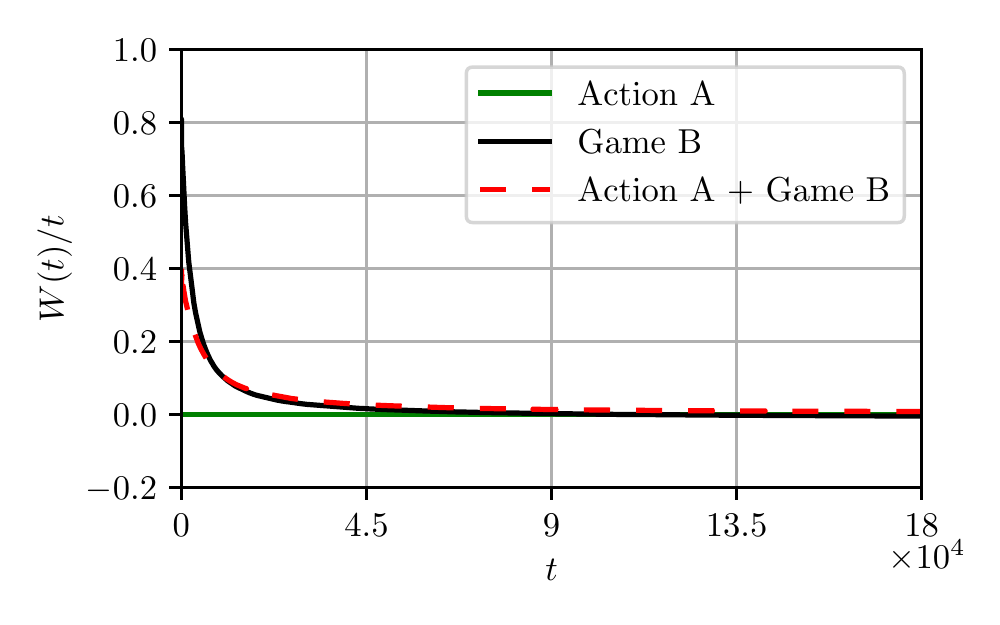}
        \cr\\[-1.8em]
        \centering{\hspace{18pt}\scriptsize(b)}
        &
        \centering{\hspace{22pt}\scriptsize(c)}
    \end{tabularx}
    \caption{(a) Simulation results of \textbf{Action A} only, \textbf{Game B} only, and the stochastically mixed \textbf{Action A} + \textbf{Game B}, across $p_1$--$p_2$ probability space. Blue highlights indicate regions where the \textit{Parrondo-type paradox} occur. The presented data were averaged over $30$ trials. (b) Change in total population gain $W(t)$ as time progresses, with $p_1=0.885$ and $p_2=0.010$. (c) Change in $W(t)/t$ as time progresses, for the 
    same parameter combination. The presented data were obtained as an average over 100 trials.}
    \label{fig:3}
\end{figure*}

Figure \ref{fig:3}(a) presents simulation results with pure \actionA, pure \gameB, and stochastically mixed \actionA + \gameB configurations. It is observed that \actionA alone does not result in any population gain, as is trivially expected from a passive rewiring process, whereas \gameB and \actionA + \gameB can result in both negative and positive population gains. Importantly, there exist regions in parameter space where Parrondo-type paradoxes occur---the combined \actionA + \gameB can result in enhanced population gain as compared to \gameB individually, and even produce positive gain when \gameB is losing. These parameter spaces demonstrate that the passive network rewiring process captured in \actionA can serve effectively as an agitative source to sustain ratcheting when mixed with the capital redistributive dynamics of \gameB.

As an illustrative example, we examine $p_1=0.885$ and $p_2=0.010$, whose corresponding evolution of population gain $W(t)$ is presented in Figure \ref{fig:3}(b). It is clear that \gameB individually produces a losing result over the long-term, whereas the stochastically mixed \actionA + \gameB leads to a sustained winning outcome. In addition, Figure \ref{fig:3}(c) presents the evolution of $d=W(t)/t$, indicative of the per-round population gain. As can be observed, $d$ gradually converges toward stable values. \gameB individually produces steady-state $d=-0.0035<0$, reflecting a losing game, and \actionA + \gameB produces steady-state $d=0.0090>0$, reflecting a winning game.

\subsection{Network structure evolution}\label{sec:simulations/structure}

To probe further into the underlying mechanism of this paradoxical effect, an analysis of the evolution of network structure is required. We define the \textit{average path length} $L$ of the network \cite{estrada2012structure} as the average distance between any two nodes:
\begin{equation}
L = \frac{2}{N(N-1)} \sum_{1\leq i \leq j \leq N} d_{ij},
\end{equation}
where $N$ is the network size and $d_{ij}$ denotes the distance between nodes $i$ and $j$ in the network as measured on the shortest connecting path.

Furthermore, the clustering coefficient $\mathcal{C}$ \cite{estrada2012structure} of the network is defined as the average of clustering coefficients $\mathcal{C}_i$ of node $i$, for all nodes within the network. This is expressed as
\begin{subequations}
    \begin{align}
        \mathcal{C}_i &= \frac{2\mathcal{E}_i}{k_i(k_i-1)} \\
        \mathcal{C} &= \frac{1}{N} \sum_{1\leq i \leq N} \mathcal{C}_i
    \end{align}
\end{subequations}
where $k_i$ is the degree of node $i$, and $\mathcal{E}_i$ is the number of edges among these $k_i$ neighbouring nodes.

\begin{figure*}[!t]
    \centering
    \begin{tabularx}{\textwidth}{@{} m{0.5\textwidth} m{0.5\textwidth} @{}}
        \centering\includegraphics[width=0.5\textwidth]{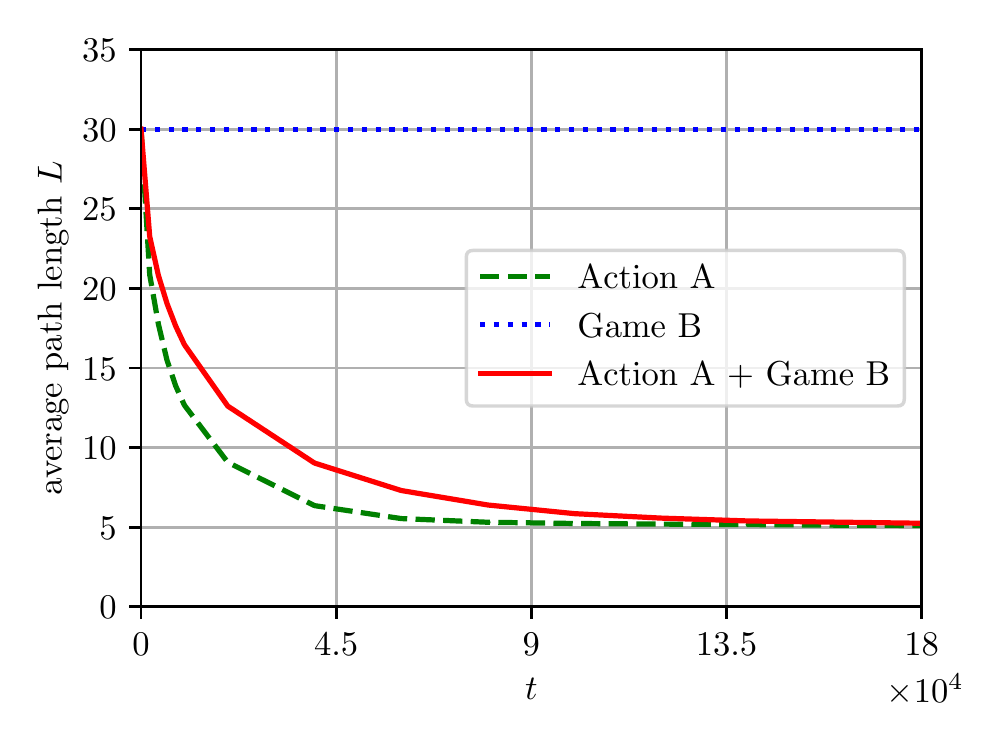}
        &
        \centering\includegraphics[width=0.5\textwidth]{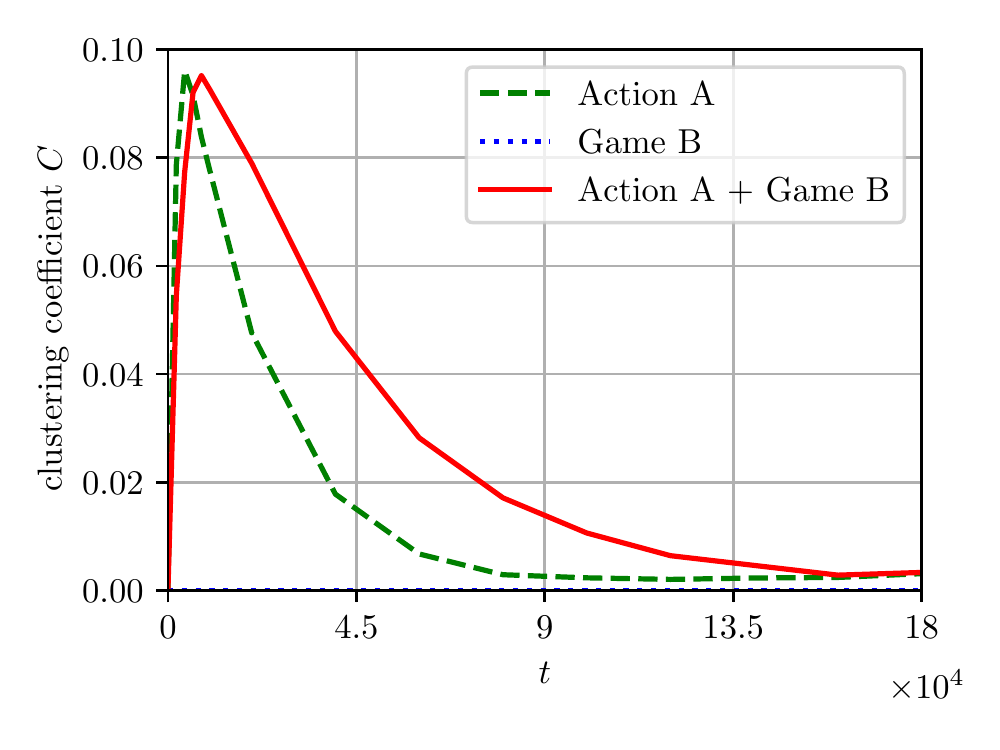}
        \cr\\[-1.8em]
        \centering{\hspace{18pt}\scriptsize(a)}
        &
        \centering{\hspace{22pt}\scriptsize(b)}
    \end{tabularx}
    \begin{tabularx}{0.6\textwidth}{@{} m{0.5\textwidth} @{}}
        \centering\includegraphics[width=0.5\textwidth]{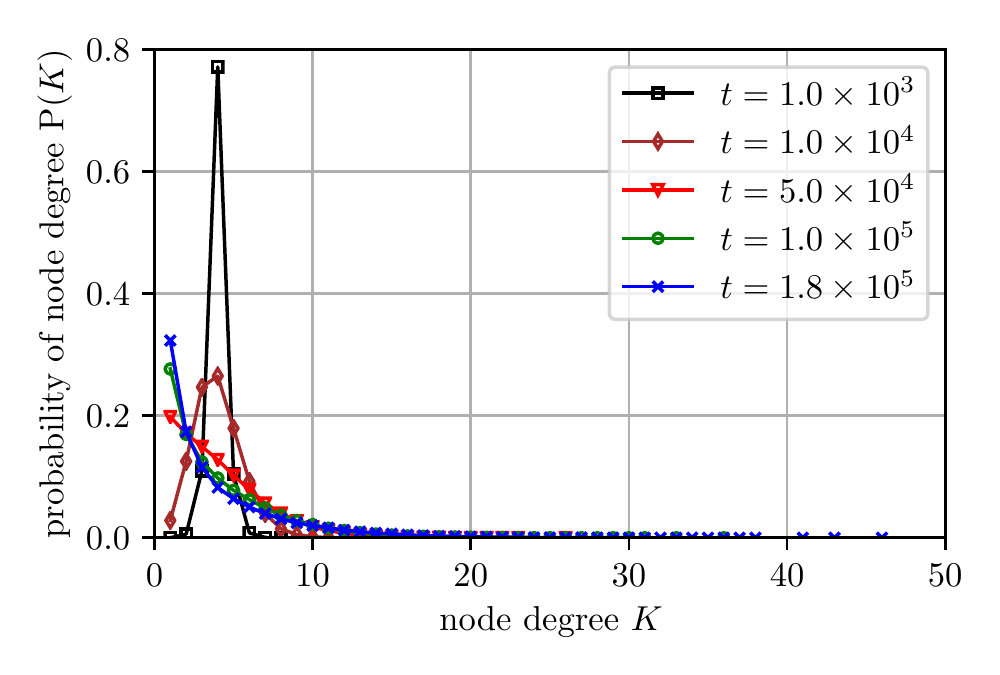}
        \cr\\[-1.5em]
        \centering{\hspace{12pt}\scriptsize(c)}
    \end{tabularx}
    \caption{Evolution of (a) average path length $L$ and (b) clustering coefficient $\mathcal{C}$ of the network for \textbf{Action A} only, \textbf{Game B} only, and stochastically mixed \textbf{Action A} + \textbf{Game B}. (c) Degree distribution of the network at varying time values as the stochastically mixed \textbf{Action A} + \textbf{Game B} progresses, illustrating a shift from initial regularity, to a normal-like distribution, and finally to a power-law distribution. Parameters are $p_1=0.885$ and $p_2=0.010$, corresponding to the demonstrative case shown previously in Figure \ref{fig:3}(b)--(c).}
    \label{fig:4}
\end{figure*}

The time-evolution of the average path length $L$ and the clustering coefficient $\mathcal{C}$ of the network are shown in Figure \ref{fig:4}(a) and \ref{fig:4}(b), under the selected parameters. The network structure is static when playing \gameB individually---the degree of all nodes in the two-dimensional lattice remains at $K=4$, and $L\approx{30}$. There are no connections between neighbors of each node, therefore $\mathcal{C}=0$. With network evolution from \actionA, there is first a rapid drop in $L$ and a rapid rise in $\mathcal{C}$. This suggests a shift towards a random or small-world topology. As evolution progresses, $L$ eventually stabilizes at $L\approx{5.2}$, whereas $\mathcal{C}$ declines after reaching a peak of $\mathcal{C}\approx{0.1}$ and eventually stabilizes at $\mathcal{C}\approx{0.004}$. The short average path length and small clustering coefficient appear consistent with scale-free networks. Furthermore, it can be observed that \actionA individually produces a faster evolution of network topology than the stochastically mixed \actionA + \gameB, as reflected by the time difference in changes of $L$ and $\mathcal{C}$. The reason for this is that in \actionA + \gameB, on average half of the elapsed game rounds are spent on \gameB, which does not entail network evolution; on the other hand the full proportion of game rounds drive network evolution in \actionA alone.

The shift in network topology is further confirmed with an analysis on the 
degree distribution of the network, as presented in Figure \ref{fig:4}(c). It 
is observed that the degree distribution morphs from the $\delta$-distribution 
of the initial lattice to a normal-like distribution as evolution progresses, 
in turn eventually shifting into a power-law distribution.  In our proposed 
rewiring process, if the degree of all neighbors of node is 1, then node is 
re-selected. To some extent, this has a "preferential attachment" effect. The 
initial network is a two-dimensional lattice network. With evolution from 
Action A, nodes with a single neighbour will gradually appear. These nodes will 
no longer be selected in subsequent rounds of network evolution. Therefore, 
with the increase of time, the set of nodes participating in Action A will 
become smaller. This leads to the formation of a tail, which may be the reason 
why the degree distribution becomes power-law over time.  Curve-fitting on the 
degree distribution data have been performed to quantitatively confirm this 
trend, with results presented in Table \ref{tab:curve-fitting}. Summarizing 
these results, under the rewiring process of \actionA, the network structure 
gradually evolves from the initial two-dimensional lattice towards a scale-free 
network. A stabilization of the network structure is observed after 
approximately $t=10^5$, as reflected in the approach towards steady-state in 
Figures \ref{fig:4}(a)--(c).

\begin{table*}[!t]
    \centering
    \caption{Curve-fitting results on network degree distribution data at varying time values, corresponding to Figure \ref{fig:4}(c).}
    \footnotesize
    \begin{tabular}[t]{l l l l}
        \toprule
        Time $t$ & Best-Fit Distribution & Function \\
        \midrule
        $1.0\times10^3$ & Normal & $P(k)=0.7718\exp{\left[-\left(\frac{k-3.993}{0.7114}\right)^2\right]}$\\
        $1.0\times10^4$ & Normal & $P(k)=0.2709\exp{\left[-\left(\frac{k-3.784}{2.061}\right)^2\right]}$\\
        $5.0\times10^4$ & Normal &
        $P(k)=0.2181\exp{\left[-\left(\frac{k+1.405}{7.279}\right)^2\right]}$\\
        $1.0\times10^5$ & Power Law & $P(k)=0.2971k^{-1.023}$\\
        $1.8\times10^5$ & Power Law & $P(k)=0.3369k^{-1.143}$\\
        \bottomrule
    \end{tabular}
    \label{tab:curve-fitting}
\end{table*}

\subsection{Macroanalysis on paradoxical mechanism}
\label{sec:simulations/macro}

The examined case of $p_1=0.885>1/2$ and $p_2=0.010<1/2$ implies that \branchOne is favourable (large winning probability) and \branchTwo in unfavourable (large losing probability). The expected gain $E$ of \gameB can be computed as
\begin{equation}
E= \pi_1 \left(2p_1-1\right) + \pi_2 \left(2p_2-1\right),
\label{eqn:mean}
\end{equation}
where $\pi_1$ and $\pi_2$ represent the stationary distribution probabilities of 
\branchOne and \branchTwo of \gameB respectively. Under $p_1=0.885$ and $p_2=0.010$, the condition for a fair \gameB ($E=0$) can be calculated to be $\pi_1^\dagger =0.56$ and $\pi_2^\dagger=0.44$.

\begin{figure}[ht!]
	\centering
	\includegraphics[width=0.5\textwidth]{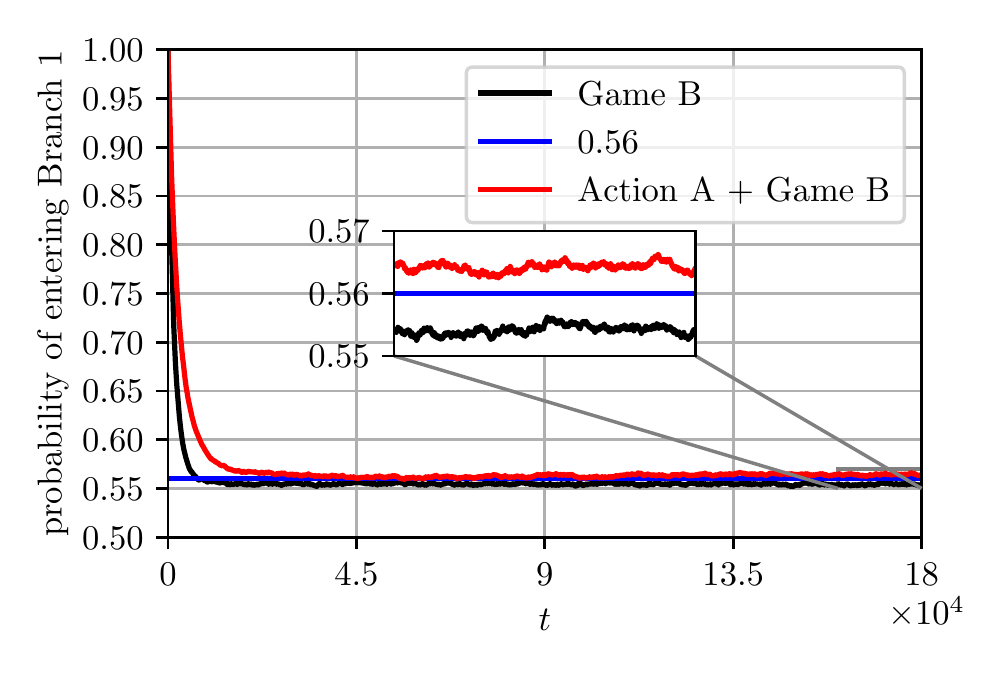}
    \caption{The probability $\pi_1$ of entering \textbf{Branch 1} in \textbf{Game B}, plotted for both a pure \textbf{Game B} sequence, and stochastically mixed \textbf{Action A} + \textbf{Game B}. The fair-game threshold of $\pi_1^\dagger=0.56$ is also plotted. Parameters are $p_1=0.885$ and $p_2=0.010$, corresponding to the demonstrative case shown previously in Figure \ref{fig:3}(b)--(c)}.
    \label{fig:5}
\end{figure}

At each game round, the average probability $\pi_1$ of playing \branchOne of \gameB can be computed by comparing the capital of each node with the mean capital of its neighbors. This is presented in Figure \ref{fig:5}. When the system settles into steady-state, $\pi_1$ stabilizes at $0.5541$ when \gameB is individually played; but this probability is raised to $0.5640$ for the stochastically mixed \actionA + \gameB. It follows from the fair-game condition that a winning outcome will result in the long-term if the steady-state $\pi_1$ exceeds $\pi_1^\dagger$, and vice versa, indeed consistent with the observation that \gameB individually is losing but \actionA + \gameB is winning. In other words, the paradoxical mechanism can be deduced to hinge upon the network structure evolution in \actionA, which raises the chance of playing the favourable branch of \gameB in subsequent rounds. This effect is large enough to overcome the losing tendency of \gameB and produce positive population gains.

\subsection{Microanalysis on positive population gains}
\label{sec:simulations/micro}

\begin{figure*}[!t]
    \centering
    \begin{tabularx}{\textwidth}{@{} m{0.5\textwidth} m{0.5\textwidth} @{}}
        \centering\includegraphics[width=0.5\textwidth]{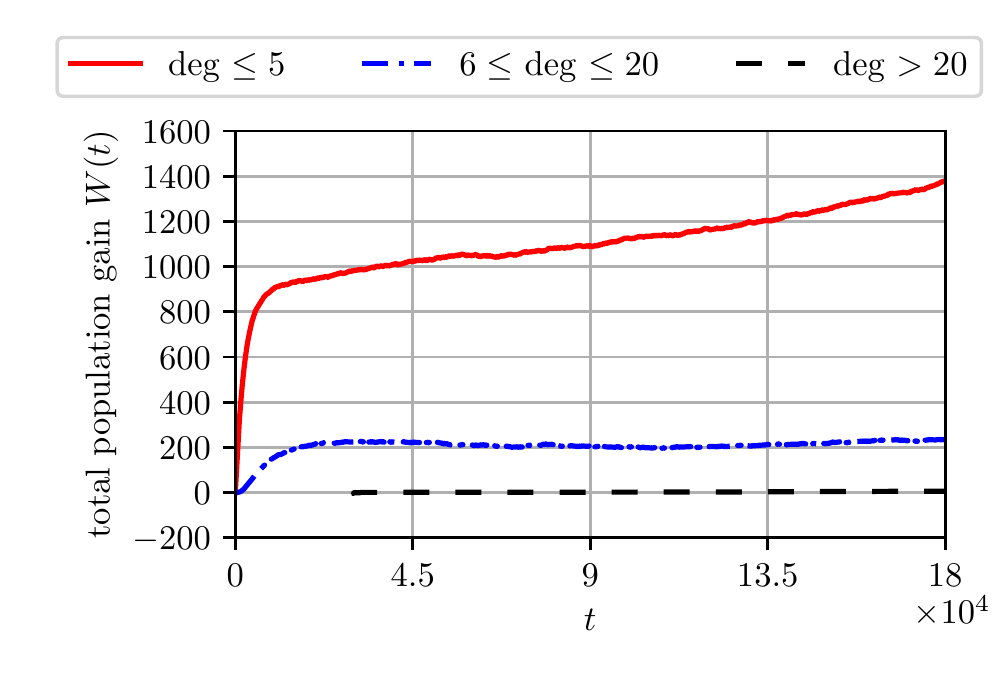}
        &
        \centering\includegraphics[width=0.5\textwidth]{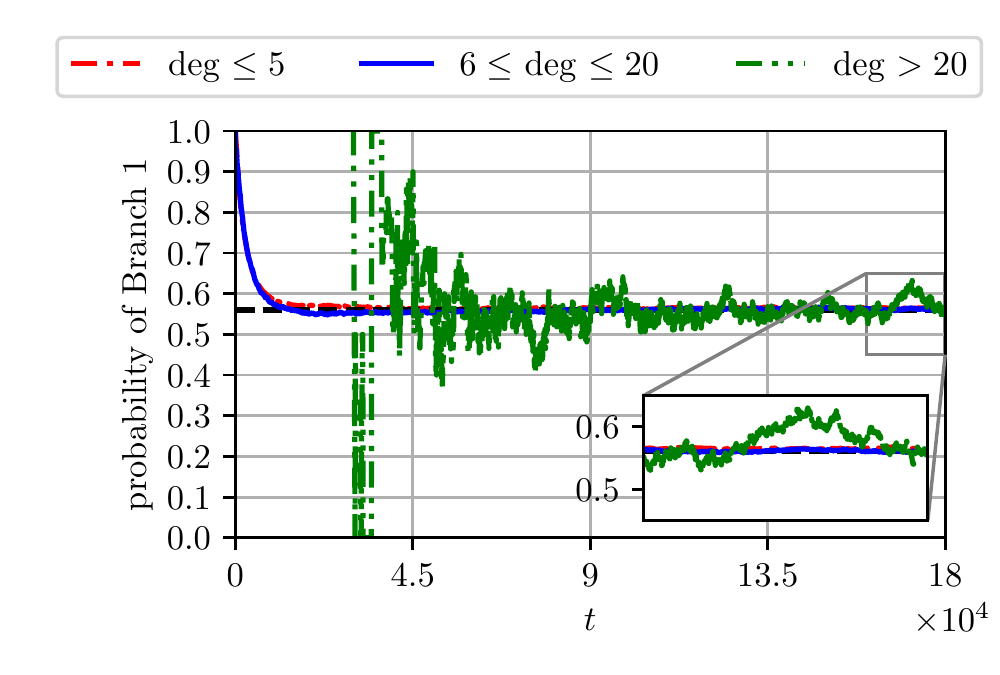}
        \cr\\[-1.8em]
        \centering{\hspace{18pt}\scriptsize(a)}
        &
        \centering{\hspace{22pt}\scriptsize(b)}
    \end{tabularx}
    \caption{(a) Total gain of the $K_i\leq{5}$, $6<K_i\leq{20}$, and $K_i>20$ subpopulations over time. (b) Plot of probability $\pi_1$ of playing \textbf{Branch 1} of \textbf{Game B} over time, for the $K_i\leq{5}$, $6<K_i\leq{20}$, and $K_i>20$ subpopulations.}
    \label{fig:6}
\end{figure*}

The criticality of network evolution in the underlying ratcheting mechanism has been suggested by the previous analysis, but the adopted statistical approach does not reveal the exact pathways through which network structure influence branch probabilities and hence long-term outcomes. We therefore seek a micro-scale analysis. To facilitate discussions, we divide the population gain into three contributory parts---gain of all nodes of degree $K_i\leq{5}$, gain of all nodes of $6<K_i\leq{20}$, and gain of all nodes of $K_i>{20}$. The gain in these three subpopulations over time is tracked and presented in Figure \ref{fig:6}(a), and the probability $\pi_1$ of the subpopulations playing the favourable \branchOne is shown in Figure \ref{fig:6}(b). 

It is observed in Figure Figure \ref{fig:6}(a) that the primary source of population gain (close to 90\%) is due to contributions from the $K_i\leq{5}$ subpopulation, followed by the $6<K_i\leq{20}$ subpopulation, with the $K_i>{20}$ subpopulation essentially negligible. At the same time, it is observed from Figure \ref{fig:6}(b) that the probability of playing \branchOne $\pi_1$ in the $K_i\leq{5}$ and $6<K_i\leq{20}$ subpopulations is consistently greater than the fair-game threshold of $0.56$, thus indicating the steady generation of positive gain. In comparison, $\pi_1$ in the $K_i>{20}$ subpopulation exhibits great volatility, suggesting rapid switching of the neighboring environments of these large-degree nodes between favorable and unfavorable conditions.

\begin{figure*}[!t]
    \centering
    \begin{tabularx}{\textwidth}{@{} m{0.5\textwidth} m{0.5\textwidth} @{}}
        \centering\includegraphics[width=0.5\textwidth]{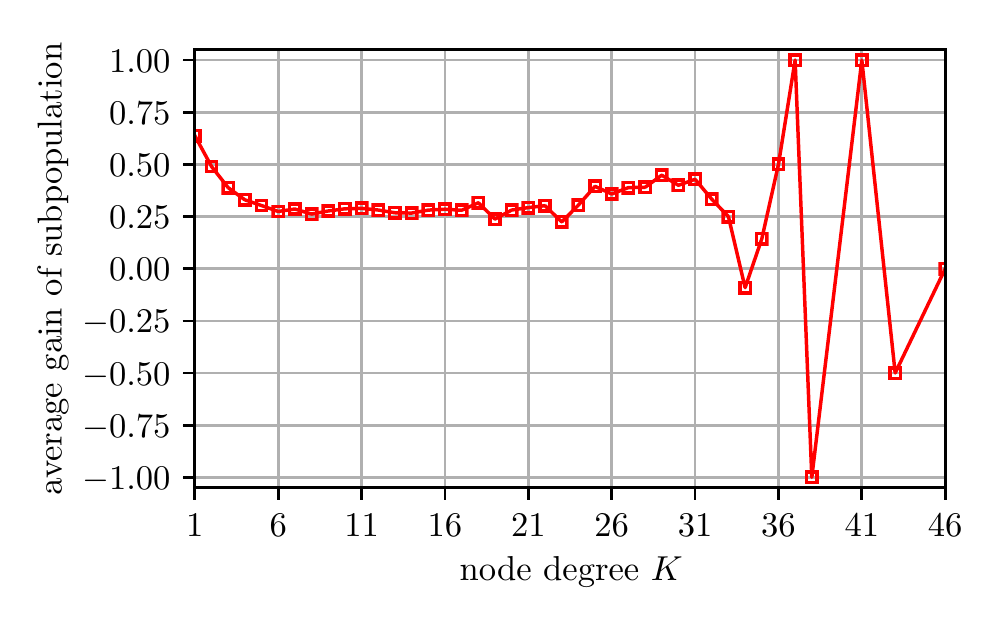}
        &
        \centering\includegraphics[width=0.5\textwidth]{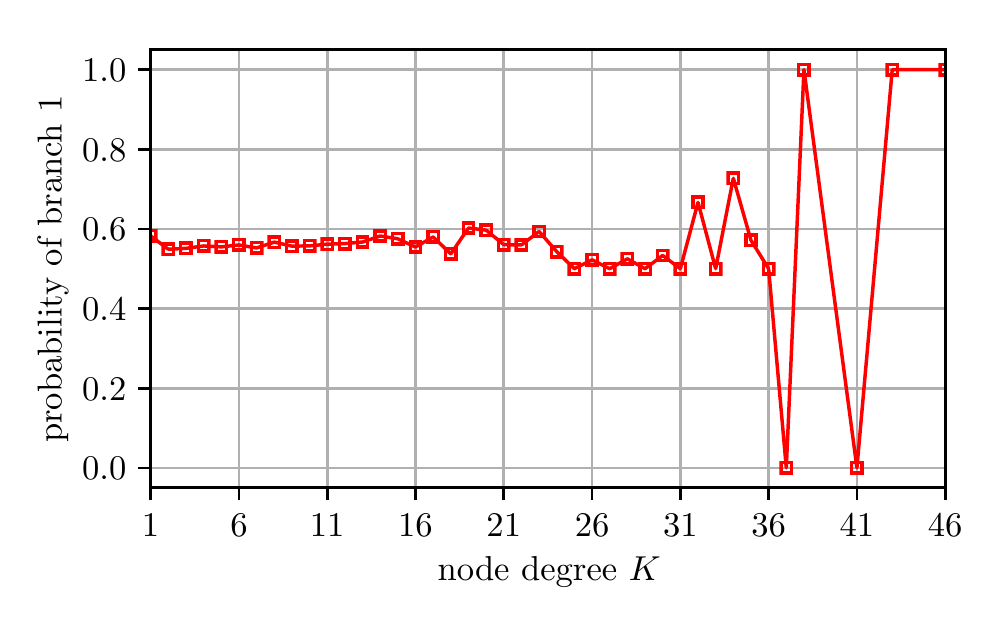}
        \cr\\[-1.8em]
        \centering{\hspace{18pt}\scriptsize(a)}
        &
        \centering{\hspace{22pt}\scriptsize(b)}
    \end{tabularx}
    \caption{(a) Steady-state average gain of subpopulations of differing node degree. (b) Steady-state probability $\pi_1$ of playing \textbf{Branch 1} of \textbf{Game B} for subpopulations of differing node degree.}
    \label{fig:7}
\end{figure*}

\begin{figure*}[!t]
    \centering
    \begin{tabularx}{\textwidth}{@{} m{0.5\textwidth} m{0.5\textwidth} @{}}
        \centering\includegraphics[width=0.5\textwidth]{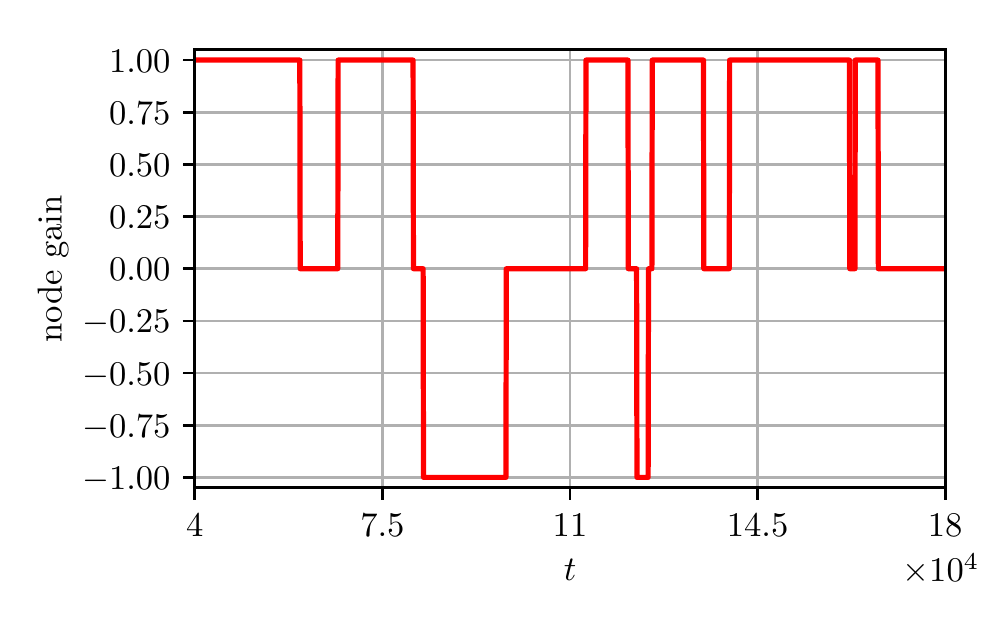}
        &
        \centering\includegraphics[width=0.5\textwidth]{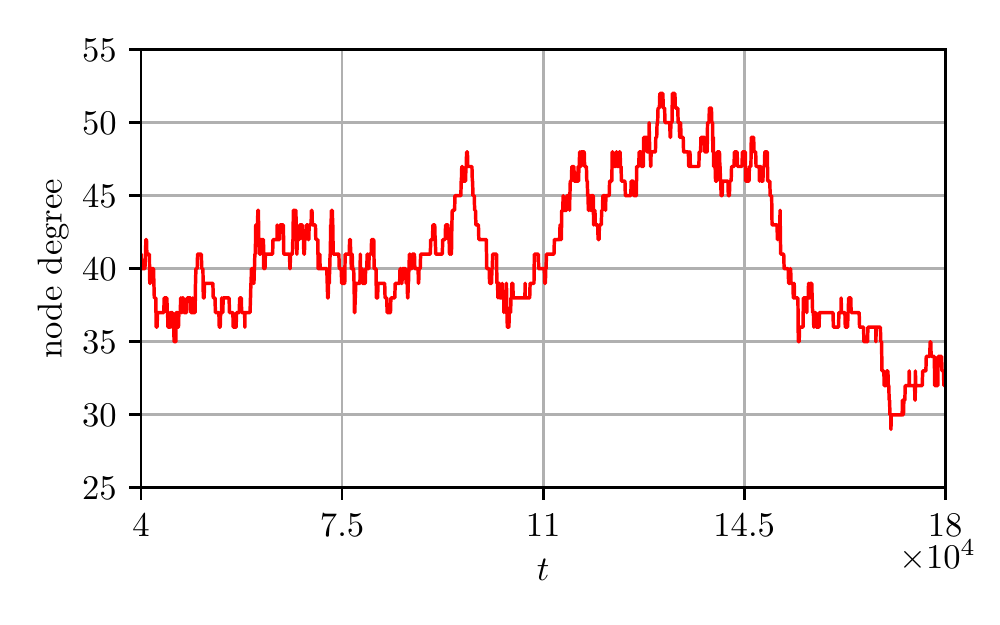}
        \cr\\[-1.8em]
        \centering{\hspace{18pt}\scriptsize(a)}
        &
        \centering{\hspace{22pt}\scriptsize(b)}
    \end{tabularx}
    \caption{(a) Evolution of the gain of a typical large-degree node over time, showing characteristic oscillatory patterns. (b) Evolution of the degree of a typical large-degree node over time.}
    \label{fig:8}
\end{figure*}

At steady-state, the average gains of subpopulations of each node degree, and their probabilities $\pi_1$ of playing \branchOne, can be computed---these results are presented in Figure \ref{fig:7}(a) and \ref{fig:7}(b) respectively. Figure \ref{fig:7}(a) confirms that the overall population gain results largely from small- and medium-degree nodes, and reveals significant polarization on large-degree nodes with alternating positive and negative average gains, averaging to negligible levels as had previously been observed. This polarization is similarly reflected in Figure \ref{fig:7}(b). Essentially, a portion of large-degree nodes have neighbouring environments that are greatly favourable in inducing \branchOne, and a portion have environments greatly unfavourable, leading to the large spread in $\pi_1$ and gain. It can, furthermore, be seen that during the evolution towards steady-state, the gain and degree of each large-degree node display large-amplitude oscillatory characteristics. This is shown in Figure \ref{fig:8}. 

These results indicate a number of key facets in the paradoxical mechanism---there is a characteristic oscillatory pattern in the gain of large-degree nodes, there is an on-average alteration of the probability of entering the favourable \branchOne, and the rewiring process of \actionA is responsible for producing paradoxical positive gains from an originally losing game, with the accompanying shift in degree distribution potentially playing a notable role. The underlying micro-scale mechanism may be summarized as follows:
\begin{itemize}
	\item[(1)] Each large-degree node is typically connected to a large number of small- and medium-degree neighbors. Suppose an arbitrary large-degree node has gain $E$ at a certain time. A large $E$ is beneficial to the neighbors of the large-degree node, as it raises their chances of playing the favourable \branchOne under \gameB. However, the resultant growth in capital of its neighbors is unfavourable to the large-degree node itself, as an increase in mean neighbor capital lowers its chances of playing \branchOne. The gain $E$ of the large-degree node will hence be eventually reduced.
	
	\item[(2)] These mechanics essentially mandate an inverse relationship between the gains of large-degree nodes and its neighbors. At some point, $E$ will have decreased sufficiently to become unfavourable to the neighbors, and the resulting decrease in mean capital of the neighbors is in turn favourable to the large-degree node. The gain $E$ will then increase, and the cycle repeats. In other words, the adversity between the large-degree node and its neighbors facilitates reversal behavior as the game progresses. This explains the alternating characteristics of the gain and $\pi_1$ of large-degree nodes previously observed in Figures \ref{fig:7} to \ref{fig:8}.
\end{itemize} 

\subsection{Sustained necessity of network rewiring agitation}
\label{sec:simulations/agitation}

\begin{figure}[ht!]
    \subfigure[]{\includegraphics{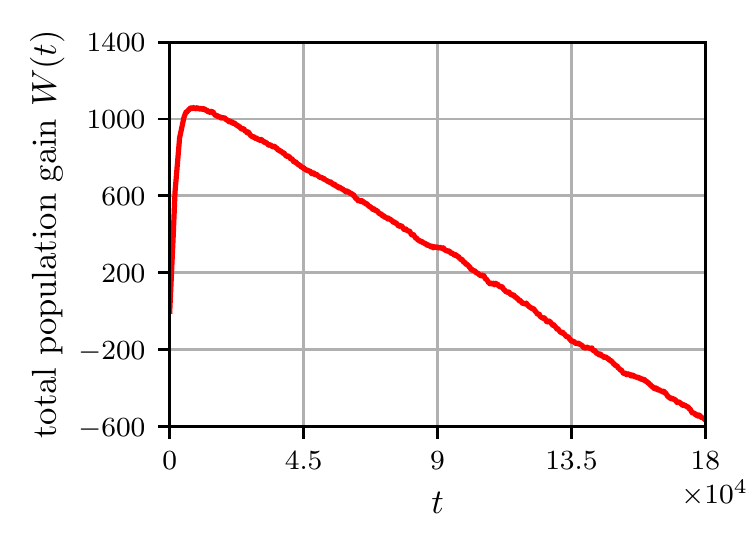}
        \label{fig:13a}} 
    \subfigure[]{\includegraphics{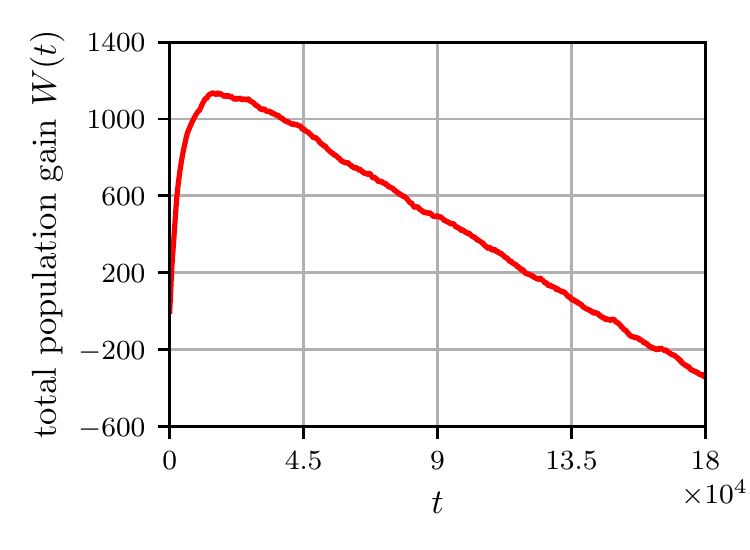}
        \label{fig:13b}}\\
    \subfigure[]{\includegraphics{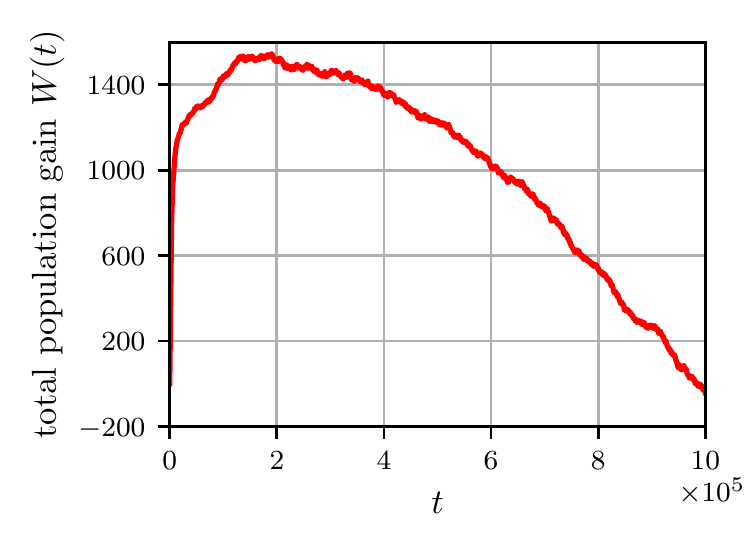}
        \label{fig:13c}}
    \subfigure[]{\includegraphics{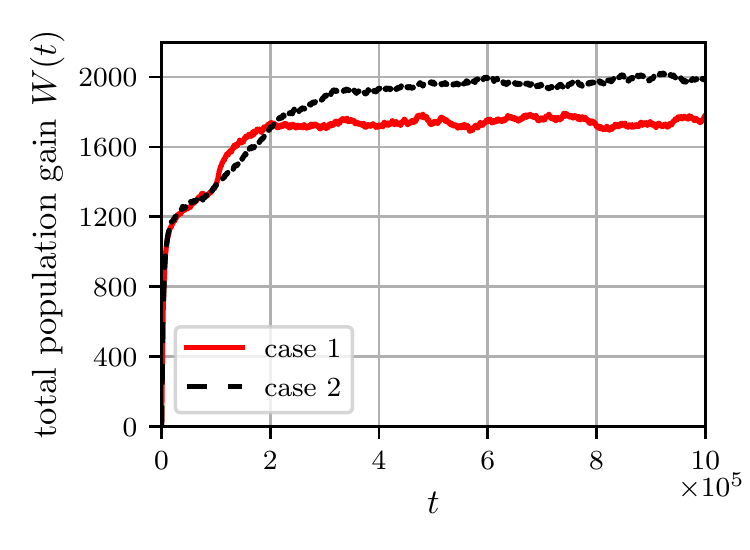}
        \label{fig:13d}}
    \caption{Plots of total population gain against time when the agitative \textbf{Action A} is omitted after a period of time $t^*$, for (a) $t^*=10^3$, (b) $t^*=10^4$, (c) $8\times 10^4$, and (d) $10^5$. In (d), simulation results for both the omission of \textbf{Action A} after $t^*$ (case 1), and the sustained progression of \textbf{Action A} + \textbf{Game B} (case 2), are shown for comparative purposes.}
    \label{fig:9}
\end{figure}

Considering the rewiring-driven mechanism inducing paradoxical population gains, it is meaningful to question if continued rewiring must occur to maintain positive outcomes, or if beyond a certain threshold, rewiring can be removed with no resultant regression. The latter may hypothetically be possible, if the initial period of rewiring is sufficient to bring the network towards a suitable topology for small-degree nodes to benefit effectively from the oscillatory patterns of larger-degree neighbors.

To investigate such a possibility, we omit \actionA after the stochastically mixed \actionA + \gameB has been played for a certain period of time $t^*$, such that \gameB is played individually for the remaining duration. The simulation results are presented in Figure \ref{fig:9}(a)--(d), for various values of $t^*$. It is observed from Figures \ref{fig:9}(a)--(c) that the omission of \actionA at too early a duration prevents the system from sustaining positive population gains, as the network topology is insufficiently evolved before entering steady-state. On the other hand,Figure \ref{fig:9}(d) demonstrates that when $t^*$ is sufficiently large ($\gtrsim{10^5}$), the system can still generate and retain a positive population gain. 

This indicates that beyond a certain period of agitation via network rewiring, the attained steady-state network structure should rewiring be disabled can confer adequate environments for small-degree nodes to sustain the exploitative mechanism. It can also be noted from Figure \ref{fig:4} that beyond $t=10^5$, the scale-free characteristics of the network structure are stable, indeed supporting this deduction. Even if rewiring is continued, Figure \ref{fig:9}(d) shows rapidly diminishing returns, and beyond $t=3\times10^5$, the agitation effect of \actionA can no longer render a growth in population gains. There is therefore a matching value between the required duration of agitative rewiring (\actionA) and the network size, exceeding which the system effectively becomes passive and insignificant further growth can be achieved.

\section{Conclusion}
\label{sec:conclusion}

In this paper, a stochastically mixed model of \actionA + \gameB was developed to study the effects of network co-evolution on group behavior adaptability and welfare dynamics. In the constructed model, \actionA reflects a rewiring process that drives evolution of network structure, and hence the survival environment of individuals, while \gameB reflects survival mechanics between interacting individuals. The influence of the environment on the survivability of individuals typically comprise both favourable and unfavourable facets, hence \gameB was constructed with two branches, of which \branchOne reflects favorable influence and \branchTwo reflects adverse influence.

The key result of the current study is that \textit{Parrondo-type} paradoxes can occur when network rewiring (\actionA) is stochastically mixed with multi-agent redistributive mechanics (\gameB)---that is, mixed \actionA + \gameB can produce long-term enhanced and even winning outcomes, despite \gameB being losing individually. It is notable that this result exhibits fundamental differences to the canonical paradoxes produced by Parrondo-type game pairs. In these canonical paradoxes, winning outcomes are produced from the interaction of two active played games (\gameA and \gameB), whereas \actionA examined here is a purely \textit{passive} process, independent of the capital of agents. The result, in other words, is that interaction between a passive process and a single game can produce paradoxes.

Through an extensive analysis, the mechanism underlying the occurrence of such paradoxes has been elucidated. Simplistically, rewiring operations shuffle connections between small-degree and large-degree nodes, thereby enabling small-degree nodes to have a greater on-average chance of invoking the favourable \branchOne through exploitation of oscillatory behavior of their large-degree neighbors. The evolution of the network structure from the initial lattice topology towards small-world and scale-free networks further facilitates such a mechanism. It is also revealed that rewiring need not be continued indefinitely for the system to maintain positive outcomes, as beyond a threshold duration, the network structure can become adequately conducive to sustain the paradoxical mechanism.

The common emergence of multi-agent competitive-cooperative behavior and interaction structure has indeed been reported in prior literature, typically analyzed through game-theoretic frameworks. Game benefits generated by competitive and cooperative relationships among individuals oftentimes indicate a degree of group adaptability. A variety of game models and contexts have been examined in these existing works, including the prisoner's dilemma model \cite{szolnoki2008towards,vukov2006cooperation}, the snowdrift game \cite{hauert2004spatial,shang2006cooperative}, the public goods game \cite{hauert2002volunteering,santos2008social}, and the ultimatum game 
\cite{nowak2000fairness,ye2016effect}, reflecting differing competitive and cooperative dynamics between individuals. Our current results will aid in the understanding of individual rationality and mobility in their communities, and the group advantages they can potentially confer. In this respect, an ecologically-relevant example is that of emperor penguins, which exhibits coordinated group behavior for survival in cold environmental conditions. Emperor penguins typically cluster in close proximity to conserve heat, but individuals in the outer ranks lose heat more quickly, and are at risk of freezing. If the outer ranks fall, the entire formation will crumple layer by layer. This is detrimental to the entire group. The penguins therefore adopt a dynamic 'transposition' group behavior, similar to the rewiring process of \actionA, that exchanges individuals between inner and outer ranks. Through this passive process, the risks of outer ranks freezing and inner ranks overheating are reduced.

The current study has focused on strictly passive rewiring mechanisms (independent of agent capitals), as such a regime is greatly complementary to the current understanding of Parrondo-type frameworks. In particular, the evolution of background dependencies between agents in the real-world is seldom tied to direct changes in capital---for instance, changes in the social network of an individual does not immediately alter career and financial standing, but may nonetheless eventually influence progression indirectly. A passive mechanism is needed to model such phenomena. Variants of capital-dependent rewiring, such as preferential rewiring towards high-capital or low-capital nodes, can be investigated in future studies.

\section*{Acknowledgments}
%This project was supported by the National Natural Science Foundation of China (Grant No.11705002); Ministry of Education, Humanities and Social Sciences research projects 15YJCZH210).
This project was supported by the National Natural Science Foundation of China (Grant No.11705002), the Ministry of Education, Humanities and Social Sciences (Research Project 15YJCZ
H210; 19YJAZH098), SUTD Start-up Research Grant (No. SRG SCI 2019 142), and the SUTD-MIT IDC Grant (No. IDG21900101 and IDIN19001).

\section*{References}

\setlength{\bibsep}{0pt}
\bibliographystyle{elsarticle-num}
\bibliography{group_welfare}

\end{document}